\documentclass[a4paper,11pt,top=1.5in,bottom=1.5in,right=1.5in,left=1.5in]{article}
\usepackage{latexsym}
\usepackage{amsmath, amsthm, amssymb}
\usepackage{graphicx}
\usepackage[Table,xcdraw]{xcolor}
\usepackage{float}
\usepackage{enumerate}
\usepackage{gensymb}
\usepackage{pdfpages}
\usepackage{comment}
\usepackage{caption}
\usepackage{subcaption}
\usepackage{boldline}
\usepackage{hyperref}
\usepackage{cite}
\usepackage{float}
\numberwithin{table}{section} \numberwithin{equation}{section}
\numberwithin{figure}{section}
\title{Polynomial Spline Regression: Theory and Application}
\author{Mithun Kumar Acharjee \\University of Alabama at Birmingham, AL\\Email: acharjee@uab.edu\\\\Kumer Das\\University of Louisiana at Lafayette, LA\\Email: kumer.das@louisiana.edu}
\date{\today} 

\begin{document}

\maketitle
\begin{abstract}
To deal with non-linear relations between the predictors and the response, we can use transformations to make the data look linear or approximately linear. In practice, however, transformation methods may be ineffective, and it may be more efficient to use flexible regression techniques that can automatically handle nonlinear behavior. One such method is the Polynomial Spline (PS) regression. Because the number of possible spline regression models is many, efficient strategies for choosing the best one are required. This study investigates the different spline regression models (Polynomial Spline based on Truncated Power, B-spline, and P-Spline) in theoretical and practical ways. We focus on the fundamental concepts as the spline regression is theoretically rich. In particular, we focus on the prediction using cross-validation (CV) rather than interpretation, as polynomial splines are challenging to interpret.  We compare different PS models based on a real data set and conclude that the P-spline model is the best. 
\end{abstract}

\parindent 0pt
\baselineskip 10pt \vskip 1pt
 
\section{Introduction}\label{intro}
Analyzing the data with linear regression models is not always sufficient. This fact has already been proven in different fields of science,especially when dealing with complex relationships such as weather data and physical relations \cite{Numerical, ENSO, nlm1, nlm2, nlm3}. This insufficiency is largely due to uncertainty about the specific form of an effect that a covariate has on the response or theoretical considerations about the application \cite{Numerical}. Variable transformations and the use of polynomials are two prevalent methods for dealing with this non-linearity. In practice, we have very few transformations. Therefore, the mathematical form of the response and the predictor variable is not only limited but also inflexible \cite{FKLM}. Furthermore, polynomial regressions need to be more flexible to capture sudden changes in slope, especially at irregular intervals. Polynomial regressions tend to fail in high dimensions due to perfect multicollinearity (since we are adding covariate squares, covariate cubes, etc.). To overcome this situation, statisticians developed a novel polynomial spline regression approach. \par

\vspace{0.2cm}
Spline regression works in two phases: the first phase ensures the partition of the domain of predictor variables into intervals, whereas the second phase estimates the polynomials separately in each interval. This two-phase procedure makes the polynomial spline model more flexible in low dimensions and less likely to generate perfect multicollinearity in higher dimensions \cite{SKP}. Spline models are theoretically vibrant, which led us to investigate different types of spline models. The main idea of the truncated polynomial (TP) based polynomial splines (PS) is to define local polynomials on intervals in the domain of the predictor variable. Also, to ensure smoothness, we need the functions to be continuous and differentiable at the interval boundaries. Despite some advantages of TP spline (such as easy understanding and simple calculation procedure), researchers are still looking for a better approach. Instead, they proposed the B-spline or basic spline, which resolves the numerical instability and co-linearity problems due to relatively closer knots. B-splines have beneficial properties, especially the derivatives of the basis function. This property has an enormous advantage in spline regression, where some significant results are discussed in this paper. Before developing the regularization-based penalty splines, one of the questions that need to be answered is: what is the number of knots? To get around this question, the P-spline approximates the function $f(z)$ with a polynomial spline that uses many knots (usually about 20–40) \cite{FKLM}. Even in the case of highly complex functions, this guarantees that $f(z)$ can be approximated with sufficient flexibility and then use the penalized least square criteria to avoid over-fitting.\par

\vspace{0.2cm}
While piecewise polynomials have been used to model regression functions for a long time \cite{hudson, fuller, studden, gallant, wold}, their application to non-parametric regression, or smoothing, is relatively new \cite{wand}. A regression spline smoothing involves transforming a regression function into a piecewise polynomial with a high number of pieces relative to the sample size. Regression spline smoothing involves modeling a regression function as a piecewise polynomial with many pieces relative to the sample size. A wide range of possible fits is possible due to the arbitrary number of polynomial pieces and the location of the knots. Because of the high number of candidate models, this flexibility poses a challenging model selection problem.\par

\vspace{0.2cm}
The study aims to achieve three specific objectives. First, we review different PS models (TP-based PS, B-spline, and P-spline) and discuss the theoretical concepts, especially the model formulation, parameter estimation, and important properties. Second, we investigate the use of these methods in modeling atmospheric pressure data, where we focused on the prediction using a more objective approach, namely cross-validation, which aims to minimize the mean square prediction error \cite{spline}. Lastly, to overcome the challenging model selection problem, we compare our models based on the six different model selection criteria and suggest the best fitted model.\par 
\vspace{0.2cm}

The rest of this article is organized into four sections: Section 2 describes the mathematical theory and properties of different spline models. Section 3 describes the data set and the steps to analyze the data. Section 4 provides the results of the analysis, while Section 5 describes and compares the findings with concluding remarks.

\section{Statistical Methods: Polynomial Spline}
The idea of the polynomial spline is closely related to polynomial regression. The polynomial regression models the covariate effects on response. A classical polynomial model can be written as:
\begin{equation}
    f(z_i)=\gamma_0+\gamma_1z_i+\dots+\gamma_l z_i^l 
\end{equation}
If we increase the degrees of the polynomial, the estimated nonlinear function $f(z)$ looks very wiggly. To make this polynomial smoother, we can partition the covariates domain into different intervals. Secondly, for each interval, we can estimate the polynomials. Thus a function $f(z)$ is a polynomial spline of degree $\textit{l}\ge 0$ with knots $a=k_1<\cdot\cdot\cdot<k_m=b$ satisfies the following two conditions \cite{FKLM}:
\begin{enumerate}
    \item $f(z)$ is continuously differentiable $(\textit{l}-1)$ times. In the case of $\textit{l}=1$, the function is only continuous (not differentiable). We do not need any smoothness requirements for $\textit{l}=0$.
    \item $f(z)$ is a polynomial of degree $\textit{l}$ on the intervals $[k_j,k_{j+1})$ defined by the knots.
\end{enumerate}
In this spline regression, the degree $\textit{l}$ indicates the overall smoothness, while the number of knots controls the number of piece-wise polynomials. There are different approaches to the polynomial spline. Some of them are discussed below.

\subsection{Polynomial spline (PS) based on truncated power (TP)}
Let the standard form of the TP-based polynomial:
\begin{equation}
y_i=\gamma_1+\gamma_2z_i+\cdot+\gamma_{l+1}z_i^l+\gamma_{l+2}(z_i-k_2)^l_+ +\gamma_{l+m-1} (z_i-k_{m-1})^l_+ + \epsilon_i
\end{equation}
 where, 
 \begin{equation}
    (z-k_j)^l_+=
    \begin{cases}
        (z-k_j)^l & \text{if } z\ge k_j\\
        0 & \text{if } z<k_j \nonumber
    \end{cases}
\end{equation}
 This model has two parts: The first part is a knot-free polynomial having l degree, which is also referred to as a global polynomial. The second part includes the knots referred to as local polynomials. In a local polynomial, the coefficient changes with the change of knots. We use this local polynomial in every interval, ensuring global smoothness. In TP-based polynomial spline, we first calculate (or draw) all the scale functions separately, and finally, we add all of them to get the estimate of $f(z)$. Thus, the final form of the TP-based PS is:
 \begin{equation}
     y_i=\sum_{j=1}^{d} \gamma_j B_j(z_j)+\epsilon_i=f(z_i)+\epsilon_i
 \end{equation}
where, \[B_1(z)=1, B_2(z)=z,\dots,B_{l+1} (z)=z^l, B_{l+2}(z)=(z-k_2)^l_+,\dots,B_d(z)=(z-k_{m-1})^l_+.\]
Here, the TP based polynomial is expressed as a linear combination of the $d=l+m-1$ functions where $m=k+1$.

It is essential to know the number and position of the knots in spline regression. There is no ``rule of thumb" for determining the number of knots. However, we can tell that if the number is high, we can get a more flexible $\hat{f(z)}$ (more wiggly) and vice-versa. On the other hand, the position of the knot is determined by different approaches, such as equidistant knots, quantile-based knot. However, these approaches cannot talk about the number of knots. This problem is solved by using the regularization of the estimation problem using penalty knots which will be discussed in section 2.3.\par
The TP basis spline functions are not bounded. As a result, calculating the TP basis functions can lead to numerical instabilities for higher values of covariates. On the other hand, if the knots are very close to each other, the TP-basis function shows high collinearity. One can solve this problem by using the basic spline or B-spline.

\subsection{B-spline:}
B-spline basis function having $\textit{l}$ degrees consists of $(\textit{l}+1)$ polynomial pieces, which are added in an (\textit{l}-1) times continuously differentiable way. The form of the basis function is: 
\begin{equation}
    f(z)=\sum_{j=1}^{d} \gamma_j B_j(z).
\end{equation}
Here, for $\textit{l}=0$, we can deduce the definition for B-spline as follows:
\begin{equation}
    B^0_j(z)=I(k_j\leq z <k_{j+1}).
\end{equation}
Where, 
 \begin{equation}
    I(k_j\leq z < k_{j+1})=
    \begin{cases}
        1 & \text{if } k_j\leq z < k_{j+1}\\
        0 & \text{if } O.W. \nonumber
    \end{cases}
\end{equation}
If we deduce the definition for $\textit{l}=1$, we get the following form \cite{Boor}:
\begin{equation}
    B^1_j(z)= \frac{z-k_{j-1}}{k_j-k_{j-1}} I(k_{j-1}\leq z <k_{j})+\frac{k_{j-1}-z}{k_{j+1}-k_j}I(k_j\leq <k_{j+1}).
\end{equation}
where the two linear segments of the basis function is \[[k_{j-1},k_j).\] and \[[k_{j},k_{j+1}).\] Similarly, the B-splines of order $\textit{l}$ (in a general expression) can be  defined recursively as:
\begin{equation}
    B^l_j(z)= \frac{z-k_{j-l}}{k_j-k_{j-l}} B^{l-1}_{j-1}(z)+\frac{k_{j+1}-z}{k_{j+1}-k_{j+1-l}}B^{l-1}_j(z).
\end{equation}
One of the critical properties of the B-spline is the derivative property. Since each basis function includes the polynomial term, it is necessary to determine its derivatives. These derivatives reveal a simple relationship between the derivative of a spline function and the B-splines of degree one less; the calculation details for this can be seen in \cite{Boor}. The derivative of the general expression form is:
\begin{equation}
   \frac{\partial}{\partial z} B^l_j (z)=l. (\frac{1}{k_j-k_{j-l}}B^{l-1}_{j-1}(z)-\frac{1}{k_{j+1}-k_{j+1-l}}B^{l-1}(z)). 
\end{equation}
Now, to get the derivative of the entire polynomial spline, we sum the equation (2.8), where the final form is as follows \cite{Boor}:
\begin{equation}
    \frac{\partial}{\partial z} \sum_{j=1}^{d} \gamma_j B_j(z_j)=l.\sum \frac{\gamma_j-\gamma_{j-1}}{k_j-k_{j-1}}B^{l-1}_{j-1}(z).
\end{equation}
This is the expression for the first-order derivatives using first-order differences of the coefficients and the basic function of the lower order \cite{Boor}. This important result will use in the P-spline in the next section.\par
Another important property of the B-spline is the unity decomposition meaning the sum of the rows of the design matrix is 1; this can be shown from the definition of the splines in \cite{Boor}. That is, \[\sum_{j=1}^{d}B_j(z)=1.\]
As a result, the design matrix has no intercept column where the expression of the design matrix is:

\[Z=\begin{bmatrix}
B^l_1(z_1) &. &B^l_d(z_1) &\\
 . & . &.&\\
 B^l_1(z_n)& . & B^l_d(z_n)& \\
\end{bmatrix}.\]

\subsection{Penalized splines (P-Splines)}
As seen in the previous subsection, the number of knots significantly impacts the estimation. In this subsection, we discuss how we can control this number using the idea of penalization. Briefly, penalization means we first use a sizeable generous number of knots, ensuring that our function is well estimated. Then, we add a penalty to avoid over-fitting and minimize a penalized least square (PLS) criterion.

\subsubsection{P-splines based on TP basis}
As seen from the model based on the TP basis equation (2.2), the $\textit{l}+1$ basis functions in this equation are just the global polynomial, and starting from the $\textit{l}+2$ term, we have the truncated polynomials which describe deviation from this polynomial. Therefore, adding a penalty on over-fitting is equivalent to adding a penalty on the corresponding truncated polynomial coefficients. This can be expressed as a penalty on this sum       $\sum_{j=l+2}^{d}\gamma_{j}^2$, which means that large coefficients are penalized. Then we estimate the coefficients by using least squares with the additional penalization term
\begin{equation}
PLS(\lambda)=\sum_{i=1}^{n}(y_i-\sum_{j=1}^{d}(\gamma_{j}B(z_i)))^2+\lambda\sum_{j=l+2}^{d}\gamma_{j}^2 
\end{equation}
where $\lambda \geq0$ is called the smoothing parameter. As $\lambda$ increases, the estimated coefficients for the truncated polynomial basis will decrease, making the plot smoother. As $\lambda$ goes to infinity, the penalized terms will go to zero, and the $f(z)$ estimate approaches an $\textit{l}$-degree polynomial. Thus, by varying $\lambda$, we can control the smoothness of the plot. One big advantage of this approach is that after a certain number of knots, our estimation will not be affected by increasing the knots if we choose $\lambda$ optimally. Therefore, this explains why this approach is to control the number of knots.

\subsubsection{P-splines based on B-splines basis:}
The penalty for truncated polynomials was evident because the basis for the global polynomial and the knot part are separate in the model formula. On the other hand, the penalty for B-splines is less obvious because the basis is formed in a way in which the global polynomial and the truncated polynomials are merged in each term \cite{spline}. Thus, we must think here from another perspective: how to smooth the overall estimated function. To characterize the smoothness of any function, using (squared) derivatives is appropriate since this represents the variability in the function. For example, we can add a penalty on the second derivative, such as
$\lambda\int(f^{''}(z))^2dz $. To do that, we can return to the derivative form of the B-spline estimated function equation (2.9). We notice from this formula that the first derivative is written in terms of the first differences of the corresponding coefficient vector; therefore, adding a penalty on the first derivative is equivalent to adding a penalty on the first differences of the coefficient vector. In general, to add a penalty based on the $r^{th}$ order derivatives, we can use differences of a higher-order $r$. This can be expressed in the penalized residual sum of squares

\begin{equation}
 PLS(\lambda)=\sum_{i=1}^{n}(y_i-\sum_{j=1}^{d}(\gamma_{j}B(z_i)))^2+\lambda\sum_{j=r+1}^{d}(\Delta^{r}\gamma_{j})^2
\end{equation}
where 
\begin{equation}
 \Delta^1\gamma_j=\gamma_{j}-\gamma_{j-1} \And \Delta^r\gamma_j=\Delta^{r-1}\gamma_{j}-\Delta^{
r-1}\gamma_{j-1} \nonumber
\end{equation}
Note here that as $\lambda$ goes to infinity, the fit approaches a polynomial of degree $r-1$ with $r^{th}$ order differences (of course, assuming $\textit{l}$). Therefore, this is one other advantage of the B-spline. In contrast to the truncated polynomial basis, which will always be of degree $\textit{l}$, the B-spline basis adds extra flexibility to choose the degree of the splines used for modeling $f(z)$ by changing the order of difference used. This provides us with additional freedom in the modeling process.

\subsubsection{Penalized least square estimation}
To solve the PLS equation, it is very convenient to write it in a matrix form because this will provide us with a general form for both TP and B-spline approaches. Also, we will see that this form works for any general penalization approach. For the case of TP basis, we can write 
\begin{equation}
    \lambda\sum_{j=l+2}^{d}\gamma_{j}^{2}=\lambda\boldsymbol{\gamma}^{'}\boldsymbol{K}\boldsymbol{\gamma} \nonumber 
\end{equation}
where $\boldsymbol{\gamma}=(\gamma_1,...,\gamma_d)^{'}\And 
\boldsymbol{K}=diag(0,...,0,1,...,1)$ with $(\textit{l}+1)$ zeros and $(m-2)$ ones. For the case of B spline, we can define difference matrices as 
\[D_1=\begin{bmatrix}
-1 & 1 & &\\
& . & . &\\
& & -1 & 1 \\
\end{bmatrix},\]
such that 
\begin{equation}
    D_1\boldsymbol{\gamma}=(\gamma_2-\gamma_1,...,\gamma_d-\gamma_{d-1})^T \nonumber
\end{equation}
and define the order $r^{th}$ difference matrix as $D_r=D_1D_{r-1}$. Then this yields the penalty
\begin{equation}
   \lambda\sum_{j=r+1}^{d}(\Delta^{r}\gamma_{j})^2=\lambda\boldsymbol{\gamma^{'}D_r^{'}D_r\gamma}=\lambda\boldsymbol{\gamma^{'}K_r\gamma} \nonumber,
\end{equation}
where $\boldsymbol{K_r}$ is called the $r^{th}$ order difference penalty matrix. Then generally, the PLS can be written as
\begin{equation}
PLS(\lambda)=(y-\boldsymbol{Z\gamma})^{'}(y-\boldsymbol{Z\gamma})+\lambda\boldsymbol{\gamma K\gamma}
\end{equation}
Therefore, the estimated coefficient vector which minimizes the PLS is
\begin{equation}
\boldsymbol{\hat{\gamma}}=(\boldsymbol{Z'Z}+\lambda\boldsymbol{K})^{-1}\boldsymbol{Z'}y.
\end{equation}
Notice that although this estimate is biased, it can be shown that if we chose $\lambda$ appropriately, the overall mean square error would be smaller in the case of the penalized estimation than in the unpenalized case. So it is a better estimator in terms of the bias-variance trade-off. Another interesting perspective of this penalization procedure can be shown if we assume the $\gamma_i$ as martingale random variables and use random walks and Bayes estimation. This will lead to the same formula of estimation. 
\subsubsection{The smoothing matrix}
Here we discuss a concept that will help us compute some important statistical properties like an estimation of error variance and confidence intervals. 
First, apply the Basis formula (TP or B-spline) to all observations and create a design matrix Z. We can write $y=Z\gamma+\epsilon$ and therefore 
\begin{equation}
\boldsymbol{\hat{y}=\hat{f(z)}=Z\hat{\gamma}}=\boldsymbol{Z}(\boldsymbol{Z'Z}+\lambda\boldsymbol{K})^{-1}\boldsymbol{Z'}y=Sy \nonumber
\end{equation}
where
\begin{equation}
S=\boldsymbol{Z}(\boldsymbol{Z'Z}+\lambda\boldsymbol{K})^{-1}\boldsymbol{Z'}
\end{equation}
is called the smoothing matrix, which we can think of as a generalization for the H matrix in linear models. To predict a future point, we can use 
\begin{equation}
\hat{f(z)}=z'(\boldsymbol{Z'Z}+\lambda\boldsymbol{K})^{-1}\boldsymbol{Z'}y=s(z)'y \nonumber,
\end{equation}
where $Z$ is the vector of basis functions evaluated at $z$. Thus, using this form $\hat{f(z)}=s(z)'y$, we can calculate the form of confidence intervals, but first, we need to assume that the errors are normally distributed.Then, we can calculate the variance of a single observation and the covariance of the joint distribution as:
\begin{equation}
Var(\hat{f(z)}=\sigma^2s(z)'s(z) \nonumber
\end{equation} and
\begin{equation}
Cov(\boldsymbol{\hat{f_r}})=\sigma^2S_rS_r' \nonumber.
\end{equation}
Using the Central limit theorem, we can calculate these confidence Intervals. \textit{Pointwise confidence intervals} with level $\alpha$ is 
\begin{equation}
\hat{f(z)}\pm z_{1-\alpha/2}\sigma\sqrt{s(z)'s(z)} \nonumber
\end{equation}
and the \textit{Simultaneous confidence bands(Bonferroni)} is 
\begin{equation}
\hat{f(z_j)}\pm z_{1-\alpha/2r}\sigma\sqrt{s(z_j)'s(z_j)} \nonumber
\end{equation}
and the \textit{Simultaneous confidence based on the joint distribution} of $(f(z_1),\dots,f(z_r))'$ is 
\begin{equation}
\hat{f(z_j)}\pm m_{1-\alpha}\sigma\sqrt{s(z_j)'s(z_j)} \nonumber
\end{equation}
where $m_{1-\alpha}$ is determined via simulation, and it defines the $1-\alpha$ quantile of the distribution of the random variable:
\begin{equation}
max_{1<j<r}|\frac{\hat{f(z_j)}-f(z_j)}{\sigma\sqrt{s(z_j)'s(z_j)}}| \nonumber
\end{equation}
In all confidence intervals, we can replace $\sigma^2$ with  $\hat{\sigma^{2}}$ for a large number of observations.
\subsubsection{Estimation of $\sigma^2$ and method of choosing $\lambda$}
To find an unbiased estimator for  $\sigma^2$, it can be shown algebraically that
\begin{equation}
E(\sum_{i=1}^{n}(y_i-\hat{f(z)}))^2=(n-tr(2S-SS'))\sigma^2+\sum_{i=1}^{n} b_i^2 \nonumber
\end{equation}
where $b_i$ is the bias of the function estimate at the point $z_i$. If we assume that the smoother is approximately unbiased, then we can define the equivalent degrees of freedom as $edf=tr(2S-SS')$, which can be approximated by
\begin{equation}
edf=tr(S),
\end{equation}
where $S$ is the smoothness matrix, then
\begin{equation}
\hat{\sigma^2}=\frac{1}{n-edf}(\sum_{i=1}^{n}(y_i-\hat{f(z)}))^2
\end{equation}
Now we can show how to choose $\lambda$; first, we should define our optimality criterion, which will be the mean square prediction error(MSE); this is the sum of the bias and variance. It follows that minimizing the mean square can be done by minimizing the cross-validation criteria. So this means choosing the $\lambda$, which minimizes 
\begin{equation}
CV=\frac{1}{n}\sum_{i=1}^{n}(y_i-\hat{f^{-i}(z_i))^2} \nonumber,
\end{equation}
where $\hat{f^{-i}(z_i)}$ denotes the estimation function when removing observation $i$, this can be shown to be equal to 
\begin{equation}
CV=\frac{1}{n}\sum_{i}^{n}(\frac{y_i-\hat{f(z_i)}}{1-s_{ii}})^2,
\end{equation}
where $s_{ii}$ are the diagonal elements of the smoothing matrix $S$, we can think of this as the leverage from the $H$ matrix in linear modeling. This can be justified theoretically by the fact that the expected value of CV is the mean square prediction error:
\begin{equation}
E(CV) \approx \frac{1}{n}\sum_{i=1}^{n}(y_{n+i}-\hat{f(z_i)})^2 \nonumber,
\end{equation}
where $y_{n+i}$ is a new observation at point $z_{i}$. Another method to choose $\lambda$ is to use the AIC criterion where the generalized AIC equation, in this case, can be shown to be 
\begin{equation}
AIC=nlog(\hat{\sigma^2})+2(edf+1)
\end{equation}
where the edf replaces the number of parameters in the AIC form in the unpenalized case. The third way is to estimate $\lambda$ as a parameter inside the model; this can be seen from the Bayesian point of view (discussed in the technical part) because from the Bayesian perspective, $\lambda=\frac{\sigma^2}{\tau^2}$, where $\tau^2$ is the prior variance assumed for the random walk. Thus, this can be treated as a parameter inside a mixed model and use the estimated procedures of mixed models to estimate both $\sigma^2$ and $\tau^2$ \cite{Bayesian}.

\section{Data and Methods}
The data are monthly averaged atmospheric pressure differences between Easter Island and Darwin, Australia. The data was collected for 168 months from both islands. Here, the response variable is the monthly averaged atmospheric pressure differences between Easter Island and Darwin, Australia. The predictor variable is the vector of time values. The data is collected from the NIST Standard Reference Database 140 \cite{ENSO}. It has been analyzed before using periodic functions, and the reported result for the sum square errors was 788 \cite{ENSO}. \par
First, we checked the linearity of the data using a residual versus fitted plot and added the ``loess" curve, see Figure 1. Exploratory data analysis suggested that the data was nonlinear. Then we fitted the different PS models. Because PS models are complicated to interpret, we focused here on the prediction. Thus, we performed a cross-validation method by partitioning our data into two groups: the training set (80\% observations) and the test set (20\% observations). We performed three PS models (TP-based PS, B-spline, and P-spline based on B-spline) for different degrees, knots, and orders of derivatives for smoothing. We applied all these models to the training data. We recorded all the comparison criteria, such as AIC, BIC, $R^2$, PRESS residuals, and cross-validated mean squared prediction error (CVMSPE). For the P-spline, a more stable and numerically preferable Generalized cross-validation criteria was used as suggested in the R function ``gam" in the \textit{mgcv} library [6]. Later, we apply the models in the test data to understand how well the models predict based on the Mean Square prediction Error (MSE) values. We choose the best model from each set of polynomial spline models (one from TP-based PS, one from B-spline, and one from P-Spline). We compare these three models based on the abovementioned criteria and select the best PS model. It is also important to note that all the analyses are carried out using an equidistant knot approach. We perform the data analysis using R \cite{rpackage}.

\section{Results}
\subsection{Detecting non-linearity}

\begin{figure}[H]
	\centering
	\includegraphics[width=1.05\linewidth]{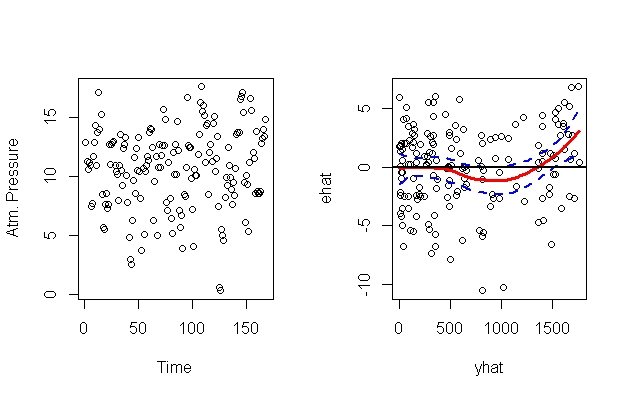}
	\caption{Detecting non-linearity: Atm. Pressure vs. Time plot (left panel) and estimated residual vs. fitted values (right panel)}
	\label{fig:Scatter Plots}
\end{figure}

\par
From Figure \ref{fig:Scatter Plots} we can tell that the data are very scattered, which shows the non-linearity pattern between the atm. pressure and time variable. The right graph indicates that the loess curve (red line) is more precise at the beginning, but after $\hat{Y}=500$, the predicted atmospheric pressure, the loess curve is less precise. Also, the Loess curve does not coincide with the horizontal error line, indicating that a non-linearity pattern exists.\par

\subsection{Model comparison}
We experimented with three models using a different number of knots, different degrees of polynomials, and different orders of differentiation (this is applicable only for P-spline). We compare the models based on different criteria. The detailed results are given in Tables  \ref{tab:TP based model}, \ref{tab:B-spline based model} and \ref{tab:P-spline based model} (see Appendix).

Based on six different model selection criteria, we choose our best model from the three different PS models shown in Tables \ref{tab:TP based model}, \ref{tab:B-spline based model} and \ref{tab:P-spline based model}. The following table shows the best model from the three methods used.

\begin{table}[H]
	\centering
	\caption{Comparing models based on the different model selection criteria}
\scalebox{0.7}{
\begin{tabular}{|c|c|c|c|c|c|c|c|c|c|}	
\hline 
	PS Models & knots& degrees & Diff.Order& AIC & BIC & $R^2$ & PRESS resid. & CVMSPE(Train.)& MSE (Test) \\ 
	\hline 
   PSM based on TP & 2 & 41 & --- & 206 & 337 & 0.98 & 850 & 6.97 & 9 \\ 
\hline
   B-spline & 2 & 41 & --- & 248 & 378 & 0.97 & 776& 6.9& 40 \\ 
    \hline
   P-spline & 2 & 80 & 1 & 186 & 343 & 0.76 & 654 & 4.9 & 8.6 \\
	\hline 
\end{tabular}
}
\label{tab: Comparing Models}
\end{table}

Table \ref{tab: Comparing Models} indicates that the AIC, Press Residuals, CVMSPE, and MSE are the lowest for P-spline regression. Though the $R^2$ is not high compared to the other model, we consider here P-spline as the best model. One possible explanation is that our focus is on the prediction, and we rely more on the values of CVMSPE and MSE.\par

\begin{figure}[H]
	\centering
	\includegraphics[width=1.05\linewidth]{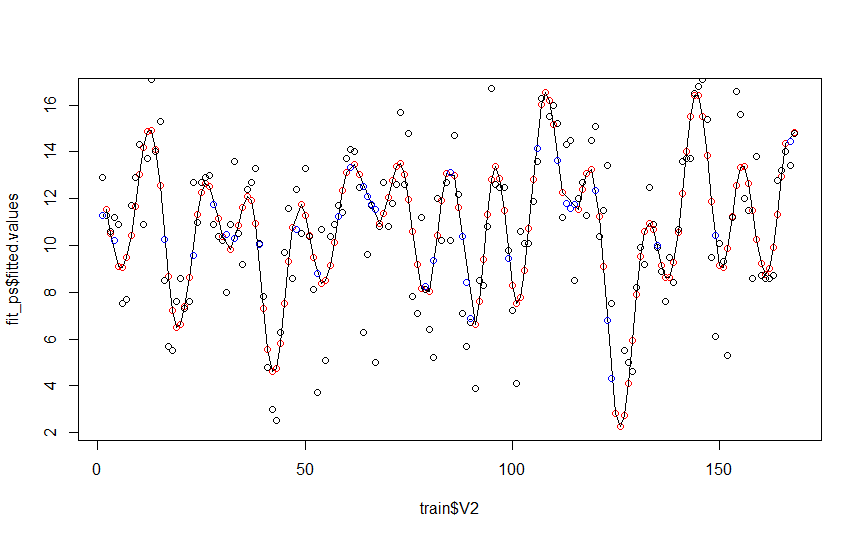}
	\caption{Plot of the best fitted model (P-spline) based on the training data (red points) and test data (blue points)}
	\label{fig: best fit}
\end{figure}

Figure \ref{fig: best fit} shows the fit of the best-fitted model (P-spline model) based on the training data set (red points) and test data set (blue points). It shows that the overall fit follows the pattern in the data and covers most of the points.

\begin{figure}[H]
	\centering
	\includegraphics[width=1.05\linewidth]{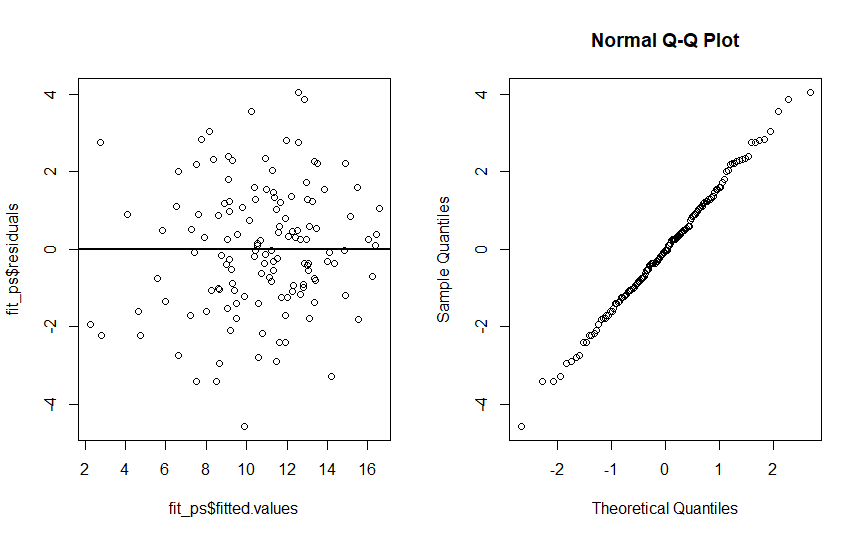}
	\caption{Residual vs Fitted graph for the best-fitted model (P-spline model)}
	\label{fig: resid vs fitted}
\end{figure}

Figure \ref{fig: resid vs fitted} shows the residual vs. fitted plot for the best model. Here the errors have no usual pattern meaning that errors are homoscedastic and are approximately independent. From here, we can assume that errors are approximately normally distributed, which will help us to construct a confidence interval based on the normality assumptions.

\section{Conclusion}
Overall, the second-degree polynomial gives better results in these three methods. In addition, since the data are very scattered, the fit improves as the number of knots increases. However, when we used P-splines, the fit tended to saturate after the number of knots exceeded 80, which is in line with the theory that the fit is not affected by the number of knots after some point. Also, it is essential to note that this data has a periodic trend, so if we try to predict a future value, we should consider this by repeating this fit periodically rather than just extending the polynomial after the last knot. This suggests that, for future improvements, periodic functions can be used with PS methods.


\section{Appendix}
\begin{figure}[H]
	\centering
	\includegraphics[width=1.05\linewidth]{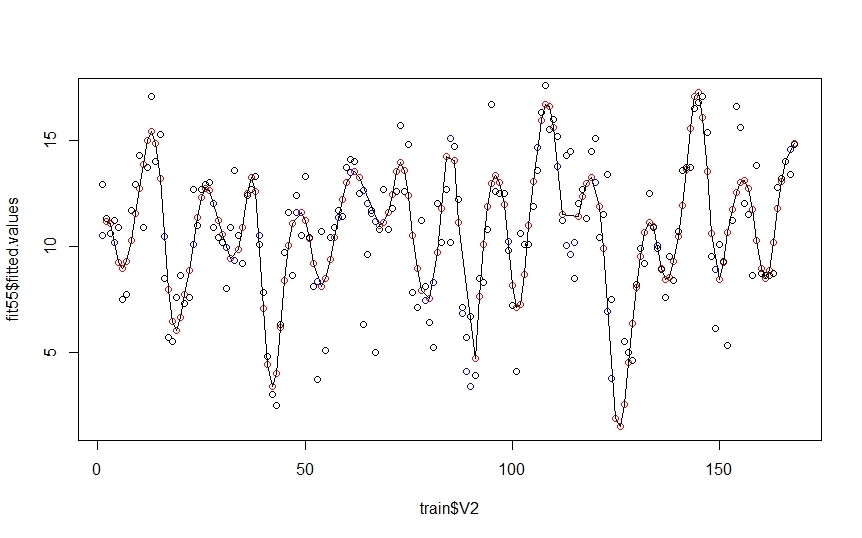}
	\caption{Best fitted model for B-spline}
	\label{fig: resid vs fitted b-spline}
\end{figure}
Figure \ref{fig: resid vs fitted b-spline} shows the fit of the best-fitted model (B-spline model) based on the training data set (red points) and test data set (blue points). It shows that the overall fit follows the pattern in the data and covers most of the points.
\begin{figure}[H]
	\centering
	\includegraphics[width=1.05\linewidth]{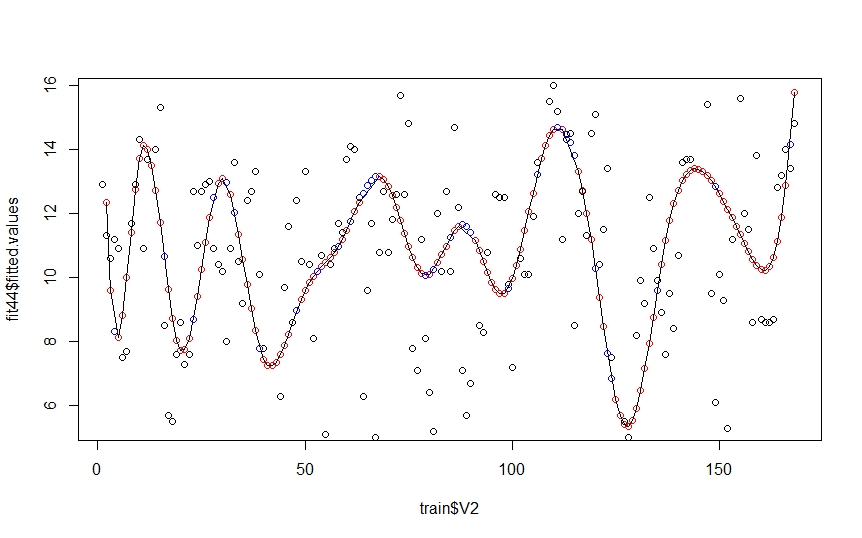}
	\caption{Best fitted model for TP based Polynomial Spline}
	\label{fig: resid vs fitted tp based spline}
\end{figure}
Figure \ref{fig: resid vs fitted tp based spline} shows the fit of the best-fitted model (TP based polynomial spline model) based on the training data set (red points) and test data set (blue points). It shows that the overall fit follows the pattern in the data and covers most of the points.
\begin{table}[H]
	\centering
	\caption{Results for Truncated Power based Polynomial Spline:}
\scalebox{0.8}{
\begin{tabular}{|c|c|c|c|c|c|c|c|c|c|}	
\hline 
	No. & knots& degrees & Parameters& AIC & BIC & $R^2$ & PRESS resid. & CVMSPE(Train.)& MSE(Test)\\ 
\hline 
 [1,]   &   1  &   1  &   3 & 338.92 & 350.51 & 0.91 &   1649.70 & 12.44  & 10.61\\
 \hline
 [2,]   &   1  &   6   &   8 &341.47 & 367.55& 0.91 &   1676.08 & 12.68  & 13.44\\
 \hline
 [3,]   &   1  &  11  &   13 & 316.16& 356.73& 0.93 &   1406.70&  10.48&   12.43\\
 \hline
 [4,]   &   1  &  16 &    18& 312.85& 367.91& 0.94 &   1360.49&  10.00&   13.57\\
 \hline
 [5,]  &    1  &  21  &   23 & 324.50& 394.04& 0.94  &  1508.76&  11.51  & 13.65\\
 \hline
 [6,]  &    1  &  26   &  28 &289.36& 373.40& 0.96 &   1148.30&   8.48&   14.28\\
 \hline
 [7,]  &    1  &  31  &   33 & 260.30& 358.83& 0.97 &    977.12&   7.91&    9.59\\
 \hline
 [8,]  &    1 &  36  &   38 &229.92& 342.93& 0.98   &  777.03&   5.97  &  8.76\\
 \hline
 [9,]  &    1 &  41 &    43 & 247.84& 375.35& 0.97&     986.12&   6.88&   10.31\\
 \hline
[10,]   &   1   & 46   &  48 &228.41& 370.41& 0.98&    874.60 &  7.73 &  10.71\\
\hline
[11,]   &   2   &  1  &    4 &340.45& 354.94& 0.91 &  1667.65 & 12.18&   10.44\\
\hline
[12,]   &   2  &   6  &    9 &336.36& 365.34 &0.92   & 1633.88&  12.29&   12.09\\
\hline
[13,]   &   2  & 11  &   14 &322.50 &365.97& 0.93 &  1449.39&  10.73 &  12.85\\
\hline
[14,]  &    2  &  16  &  19& 316.57& 374.53& 0.94 &   1420.19 & 10.26 &  14.62\\
\hline
[15,]   &   2  &  21  &   24& 297.91& 370.35& 0.95 &   1246.65 &  9.78 &  16.09\\
\hline
[16,]   &   2  &  26  &   29& 281.73 &368.67& 0.96&    1215.20&   9.79 &  10.84\\
\hline
[17,]  &    2  &  31  &   34 &245.35& 346.78 &.97 &    896.67  & 7.01 &   8.23\\
\hline
[18,]   &   2 &   36  &   39& 226.30 & 342.21& 0.98 &    772.56 &  7.76  & 10.00\\
\hline
[19,]  &   2 &   41  &   44& 206.36& 336.76& 0.98 &    849.08&   6.76  &  9.19\\
\hline
[20,]   &   2   & 46  &   49 &222.02& 366.& 0.98 &  10136.50&  68.47  &  9.77\\
\hline
[21,]   &   3  &   1   &   5 & 338.55& 355.94& 0.91 &   1639.46&  12.53 &  11.45\\
\hline
[22,]   &   3  &   6  &   10& 345.52& 377.40& 0.91 &   1724.28&  12.84  & 13.51\\
\hline
[23,]   &   3  & 11  &   15& 318.28& 364.65 &.93  &  1439.08 & 10.86 &  12.65\\
\hline
[24,]   &   3  &  16  &   20 &306.46 &367.32& 0.94  &  1329.98 & 10.07 &  14.66\\
\hline
[25,]   &   3  &  21  &   25 &319.85 &395.19& 0.94  &  1711.94 & 12.37 &  16.01\\
\hline
[26,]   &   3  &  26  &   30 &297.35 &387.18 &0.95 &  1319.39& 121.62 &  15.78\\
\hline
[27,]   &   3  &  31  &   35& 239.48 &343.80& 0.97 &   1085.65&   8.14 &   9.73\\
\hline
[28,]  &    3 &   36  &   40 & 217.75 &336.56 &0.98  &  1620.05 &  8.21 &   8.38\\
\hline
[29,]   &   3 &   41  &   45& 225.43& 358.73& 0.98&   25348.28 & 12.09&    9.04\\
\hline
[30,]   &   3  &  46  &   50& 222.19& 369.98& 0.98& 1652263.27 & 33.49&    9.56\\
\hline
[31,]   &   4  &   1  &    6& 341.21& 361.49& 0.91&    1666.24 & 12.77&   11.14\\
\hline
[32,]   &   4  &   6  &   11& 329.88& 364.65& 0.92&    1531.11&  11.63 & 13.05\\
\hline
[33,]   &   4  &  11  &   16& 321.43& 370.69& 0.93&    1419.22 & 10.84 &  12.46\\
\hline
[34,]   &   4  &  16  &   21 &312.67& 376.43& 0.94&    1609.72 & 12.49 &  16.25\\
\hline
[35,]   &   4  & 21   &  26 &310.37& 388.61& 0.95 &   1578.73 & 12.76  & 16.30\\
\hline
[36,]   &   4  &  26  &   31& 276.18& 368.92& 0.96&    2704.06 & 15.66 &  14.16\\
\hline
[37,]   &   4  &  31  &   36& 227.09& 331.41& 0.98&    1202.93 &  8.40 &   8.37\\
\hline
[38,]   &   4  &  36  &   41& 218.30& 337.11& 0.98 &    728.89 &  5.85 &   9.71\\
\hline
[39,]   &   4  &  41  &   46&215.19 &348.49& 0.98 &  1221.64&332.95 &   8.66\\
\hline
[40,]   &   4  &  46  &   51& 221.55& 369.34& 0.98 &   4310.54&  19.04  & 10.55\\
\hline 
\end{tabular}
}
\label{tab:TP based model}
\end{table}

\begin{table}[H]
	\centering
	\caption{Results for B-Spline model:}
\scalebox{0.8}{
\begin{tabular}{|c|c|c|c|c|c|c|c|c|c|c|}	
\hline 
	No. & knots& degrees &  Parameters& AIC & BIC & $R^2$ & PRESS resid. & CVMSPE(Train.)& MSE(Test)\\ 
\hline 
 [1,]  &     1 &    1  &     2&  452.38&  461.07 & 0.78 &  3806.59 & 1.224000e+01  &  30.11\\
 \hline 
 [2,]  &     1 &     6 &      7&  394.39&  417.57 & 0.87 &  2441.86 & 1.271000e+01  &  21.66\\
 \hline 
 [3,]  &     1 &    11 &     12&  351.18&  388.85 & 0.91&   1773.43 & 1.082000e+01  &  20.86\\
 \hline 
 [4,]  &     1 &    16 &     17&  332.32&  384.49&  0.93 &  1521.53&  1.036000e+01 &   21.04\\
 \hline 
 [5,]  &     1 &    21 &     22 & 345.03&  411.68 & 0.93 &  1697.94 & 1.180000e+01 &   21.32\\
 \hline 
 [6,]  &     1 &    26 &     27&  321.12&  402.26 & 0.94 &  1378.68&  8.660000e+00 &   22.30\\
 \hline 
 [7,]  &     1 &    31 &     32&  297.91&  393.54 & 0.96 &  1177.24 & 8.280000e+00  &  17.73\\
 \hline 
 [8,]  &     1 &    36 &     37&  275.05&  385.17&  0.97 &   955.09 & 6.400000e+00  &  17.35\\
 \hline 
 [9,]  &     1 &    41 &     42 & 286.47&  411.08 & 0.97 &  1159.86 & 8.540000e+00  &  20.64\\
 \hline 
[10,]  &     1 &    46 &     47&  269.52&  408.62 & 0.97&   1021.06 & 6.020000e+00  &  24.64\\
\hline 
[11,]  &     2 &     1 &      3&  414.10&  425.69 & 0.84 &  2856.68&  1.258000e+01  &  24.50\\
\hline 
[12,]  &     2 &     6 &      8&  363.99&  390.07 & 0.90 &  1992.99&  1.215000e+01 &   20.47\\
\hline 
[13,]  &     2 &    11 &     13&  337.89&  378.46 & 0.92 &  1590.55 & 1.120000e+01 &   20.97\\
\hline 
[14,]  &     2 &    16 &     18&  333.83&  388.89&  0.93 &  1573.21 & 1.060000e+01 &   23.31\\
\hline 
[15,]  &     2 &    21 &     23&  325.11&  394.66 & 0.94 &  1466.76&  9.430000e+00 &   24.26\\
\hline 
[16,]  &     2 &    26 &     28&  310.99&  395.03&  0.95 &  1403.64 & 9.280000e+00 &   20.26\\
\hline 
[17,]  &     2 &    31 &     33&  278.18&  376.71 & 0.96 &  1061.22&  7.140000e+00 &   23.87\\
\hline 
[18,]  &     2 &    36 &     38&  262.05&  375.06&  0.97 &   893.59&  5.980000e+00 &   34.14\\
\hline 
[19,]  &     2 &    41 &     43&  248.98&  376.48 & 0.97& 776.05 & 7.830000e+00  &  40.50\\
\hline 
[20,]  &     2 &    46 &     48&  263.08&  405.08 & 0.97 &  1026.91&  3.638597e+05  &  49.43\\
\hline 
[21,]  &     3 &     1 &      4&  402.06&  416.55 & 0.85 &  2612.20&  1.248000e+01 &   20.09\\
\hline 
[22,]  &     3 &     6 &      9&  362.66&  391.64 & 0.90 &  1948.77&  1.259000e+01 &   22.20\\
\hline 
[23,]  &     3 &    11 &     14&  331.11&  374.58&  0.93 &  1536.51&  1.046000e+01 &   21.98\\
\hline 
[24,]  &     3 &    16 &     19&  330.20&  388.15&  0.93 &  1541.27 & 1.029000e+01 &   22.42\\
\hline 
[25,]  &     3 &    21 &     24 & 340.65&  413.10&  0.93 &  1881.85&  1.329000e+01 &   25.70\\
\hline 
[26,]  &     3 &    26 &     29&  316.51&  403.44&  0.95 &  1447.70 & 3.413700e+02 &   42.32\\
\hline 
[27,]  &     3 &    31 &     34&  269.04 & 370.47 & 0.97 &  1037.21 & 2.079307e+06 &   50.48\\
\hline 
[28,]  &     3 &    36 &     39&  255.13 & 371.04 & 0.97 &   904.81 & 9.950000e+00 &   58.94\\
\hline 
[29,]  &     3 &    41 &     44&  263.62 & 394.02 & 0.97 &  1245.69 & 1.815200e+02  &  75.02\\
\hline 
[30,]  &     3 &    46 &     49&  264.11 & 409.00 & 0.97 &  4462.89&  1.273000e+04 &  108.50\\
\hline 
[31,]  &     4 &     1 &      5&  387.51 & 404.90&  0.87 &  2344.45&  1.243000e+01 &   17.18\\
\hline 
[32,]  &     4 &     6 &     10&  346.74 & 378.61 & 0.91 &  1724.99&  1.159000e+01 &   22.08\\
\hline 
[33,]  &     4 &    11 &     15&  337.80&  384.16 & 0.92 &  1599.50 & 1.167000e+01 &   22.36\\
\hline 
[34,]  &     4 &    16 &     20&  334.91&  395.76 & 0.93 &  1756.78&  1.229000e+01 &   22.73\\
\hline 
[35,]  &     4 &    21 &     25&  327.08 & 402.43&  0.94 &  1724.88&  2.747000e+01 &   46.33\\
\hline 
[36,]  &     4 &    26 &     30&  296.58 & 386.41 & 0.95 &  2056.64 & 5.705241e+09 &   74.20\\
\hline 
[37,]  &     4 &    31 &     35&  261.96 & 366.28 & 0.97 &  1337.21&  1.554000e+01 &   68.58\\
\hline 
[38,]  &     4 &    36 &     40&  257.76&  376.57 & 0.97 &  1247.29 & 3.308500e+02 &   89.70\\
\hline 
[39,]  &     4 &    41 &     45&  258.28&  391.58&  0.97 &  7550.26&  9.868560e+03 &  139.80\\
\hline 
[40,]  &     4 &    46 &     50&  265.38 & 413.17 & 0.97&  24097.47&  2.303280e+03&  103.93\\
\hline
\end{tabular}
}
\label{tab:B-spline based model}
\end{table}

\begin{table}[H]
	\centering
	\caption{Results for P-Spline model:}
\scalebox{0.75}{
\begin{tabular}{|c|c|c|c|c|c|c|c|c|c|c|}	
\hline 
No. & knots& degrees & Parameters& Differ.O& AIC & BIC & $R^2$ & PRESS resid. & CVMSPE (Train)& MSE(Test) \\ 
\hline 
 [1,]   &   2  &  50 & 38.63  &      0 &215.28& 330.13 &0.68& 709.99& 5.45 &   8.19\\
 \hline
 [2,]   &   2  &  50 & 39.85 &       1 &215.74& 334.12& 0.68& 716.19& 5.51 &   8.60\\
 \hline
 [3,]  &   2  &  80 & 56.09  &      0 &182.87 &348.30& 0.76& 668.20& 4.94 &   7.65\\
 \hline
 [4,]  &    2 &   80 & 53.11 &       1 &186.24& 343.04& 0.75& 654.57& 4.91 &   8.61\\
 \hline
 [5,]  &    2  & 100 & 63.46  &      0 &181.21& 367.99& 0.77& 710.59& 5.33 &   7.88\\
 \hline
 [6,]  &    2  & 100 & 55.80 &       1 &184.43& 349.02& 0.76& 663.21& 4.98 &   8.66\\
 \hline
 [7,]  &    3 &  50 & 38.71  &      0 &205.04 &320.10 &0.70 &670.43& 5.05 &   8.61\\
 \hline
 [8,]  &    3  &  50 & 39.93 &       1 &205.99& 324.59& 0.70& 673.93& 5.12 &   8.82\\
 \hline
 [9,]  &    3 &   50 & 39.95 &       2& 206.40 &325.07& 0.70& 678.38& 5.14 &   8.67\\
 \hline 
[10,]  &    3  &  80 & 50.65 &       0& 197.52 &347.20& 0.73& 700.90& 5.22 &   8.57\\
\hline
[11,]  &    3  &  80 & 47.16 &       1& 200.17& 339.73& 0.72& 683.95& 5.17 &   8.99\\
\hline
[12,]  &    3  &  80 & 43.98 &       2& 204.16& 334.51& 0.71& 683.15& 5.20 &   8.74\\
\hline
[13,]  &    3  & 100 & 63.76 &       0 &172.66& 360.31& 0.78& 671.39& 5.02 &   8.04\\
\hline
[14,]  &    3 &  100 & 54.03 &       1 &187.64& 347.12 &0.75 &667.33& 5.01  &  8.82\\
\hline
[15,]  &    3 &  100 & 46.86 &       2& 199.57& 338.26& 0.72 &674.89& 5.13  &  8.73\\
\hline
[16,]  &    4  &  50 & 37.29 &       0& 210.91 &321.86& 0.69& 684.73& 5.23  &  8.38\\
\hline
[17,]  &    4  &  50 & 38.50 &       1& 211.55 &326.01& 0.69& 688.02& 5.29  &  8.57\\
\hline
[18,]  &    4  &  50 & 38.59 &       2 &211.50& 326.22& 0.69&691.36& 5.30  &  8.47\\
\hline
[19,]  &    4  &  50 & 38.39 &       3 &211.45 &325.59& 0.69 &690.95& 5.29  &  8.39\\
\hline
[20,]  &    4  &  80 & 51.26 &       0 &189.61 &341.04& 0.75 &664.84& 4.95  &  8.19\\
\hline
[21,]  &    4  &  80 & 47.59 &       1& 197.42& 338.23& 0.73& 669.96& 5.08  &  8.81\\
\hline
[22,]  &    4  &  80 & 43.56 &       2 &203.98& 333.12 &0.71& 678.11& 5.17  &  8.67\\
\hline
[23,]  &    4  &  80 & 40.67 &       3 &207.69& 328.45& 0.70 &685.08& 5.21  &  8.49\\
\hline
[24,]  &    4  & 100 & 57.91 &       0 &182.95& 353.67& 0.76& 681.15& 5.04  &  8.01\\
\hline
[25,]  &    4  & 100 & 51.01 &       1 &192.60& 343.31& 0.74 &670.63& 5.05  &  8.81\\
\hline
[26,]  &    4  & 100 & 45.21 &       2 &201.94& 335.84& 0.72& 677.25& 5.16  &  8.70\\
\hline
[27,]  &    4  & 100 & 41.19 &       3 &207.18 &329.45& 0.70 &685.70& 5.21  &  8.51\\
\hline
[28,]  &     5 &   50 & 37.51&        0 &207.00& 318.59& 0.69& 668.39& 5.09 &   8.42\\
\hline
[29,]  &    5  &  50 & 38.48 &       1& 208.06& 322.46& 0.69 &672.60& 5.16  &  8.50\\
\hline
[30,]  &    5  &  50 & 38.49 &       2 &208.62& 323.04& 0.69&678.55& 5.18  &  8.41\\
\hline
[31,]  &    5  &  50 & 38.29 &       3 &209.18&323.03& 0.69 &677.31& 5.19   & 8.36\\
\hline
[32,]  &    5  &  50 & 38.03 &       4 &209.91 &323.03& 0.69& 666.51& 5.22  &  8.32\\
\hline
[33,]  &    5  &  80 & 46.76 &       0& 198.83& 337.24& 0.72 &680.66& 5.10  &  8.52\\
\hline
[34,]  &    5  &  80 & 44.34 &       1& 203.17 &334.55& 0.71& 679.82& 5.17  &  8.88\\
\hline
[35,]  &    5  &  80 & 41.82 &       2& 206.57& 330.66& 0.70& 683.41& 5.21  &  8.65\\
\hline
[36,]  &    5  &  80 & 39.90 &       3& 208.65& 327.17& 0.69 &686.43& 5.23  &  8.47\\
\hline
[37,]  &    5  &  80 & 38.68 &       4& 209.96& 324.95& 0.69& 671.03& 5.24  &  8.36\\
\hline
[38,]  &    5 &  100 & 56.35 &       0 &183.76 &349.96& 0.76& 672.66& 4.99  &  8.17\\
\hline
[39,]  &    5 &  100 & 49.13 &       1 &195.71 &340.98& 0.73& 672.86& 5.08  &  8.86\\
\hline
[40,]  &    5  & 100 & 43.98 &       2 &203.67& 334.01& 0.71& 679.42& 5.18  &  8.70\\
\hline
[41,]  &    5  & 100 & 40.67  &      3 &207.79& 328.53& 0.70 &686.08& 5.22  &  8.50\\
\hline
[42,]  &    5  & 100 & 38.94  &      4 &209.71& 325.43& 0.69& 671.64& 5.24  &  8.38\\
\hline
\end{tabular}
}
\label{tab:P-spline based model}
\end{table}
\end{document}